\documentclass[usenatbib]{mn2e}
\usepackage{journals}
\usepackage{graphicx}

\usepackage{amsmath}
\usepackage{natbib}

\title{Interstellar Scintillation observations for PSR B0355+54}

\author[Xu et al.]{Y.\,H.\,Xu$^{1,3,5}$,
K.\,J.\,Lee\thanks{email:kjlee@pku.edu.cn}$^{2,7}$,
L.\,F.\,Hao$^{1,5}$,
H.\,G.\,Wang$^{4}$,
Z.\,Y.\,Liu$^{6}$,
\newauthor
Y.\,L.\,Yue$^{7}$,
J.\,P.\,Yuan$^{6}$,
Z.\,X.\,Li$^{1,3,5}$,
M.\,Wang$^{1,5}$,
J.\,Dong$^{1,5}$,
\newauthor
J.\,J.\,Tan$^{1,5}$,
W.\,Chen$^{1,5}$,
J.\,M.\,Bai$^{1,5}$\\
$^{1}${Yunnan Astronomical Observatory, Chinese Academy of Sciences,
Kunming 650011, China}\\
$^{2}${Kavli Institute for Astronomy and Astrophysics, Peking
University, Beijing 100871, China. kjlee@pku.edu.cn}\\
$^{3}${University of Chinese Academy of Sciences, Beijing 100049, China}\\
$^{4}${Center for Astrophysics, Guangzhou University, Guangzhou 510006, China}\\
$^{5}${Key Laboratory for Structure and Evolution of Celestial
Objects. Chinese Academy of Sciences, Kunming 650011, China}\\
$^{6}${Xinjiang Astronomical Observatory, Chinese Academy of Sciences,
150 Science 1-street, Urumqi, Xinjiang 830011, China}\\
$^{7}${National Astronomical Observatories, Chinese Academy of
Sciences, Beijing, China}\\
}
\begin{document}
\date{Accepted 2018 February 27. Received  2018 February 27; in original form 2017 October 2017}

\pagerange{\pageref{firstpage}--\pageref{lastpage}} \pubyear{2013}

\maketitle

\label{firstpage}

\begin{abstract} In this paper, we report our investigation of
pulsar scintillation phenomena by monitoring PSR B0355$+$54 at 2.25
GHz for three successive months using \emph{Kunming 40-m radio
telescope}. We have measured the dynamic spectrum, the two-dimensional
correlation function, and the secondary spectrum. In those observations
with high signal-to-noise ratio ($S/N\ge100$), we have detected the
scintillation arcs, which are rarely observable using such a small
telescope. The sub-microsecond scale width of the scintillation
arc indicates that the transverse scale of structures on scattering
screen is as compact as AU size.  Our monitoring has also shown
that both the scintillation bandwidth, timescale, and arc curvature
of PSR B0355$+$54 were varying temporally.  The plausible explanation
would need to invoke multiple-scattering-screen or multiple-scattering-structure
scenario that different screens or ray paths dominate the scintillation process
at different
epochs.  \end{abstract}


\section{Introduction}

There are about 2000 known pulsars in the Galaxy. Their dispersion measure
and parallax provide the distance information, which make pulsars unique probes
to study the interstellar medium (ISM). The scintillation of
pulsars (see review by \citet{N92}) is a powerful tool to
investigate the ISM fluctuation and the turbulent dynamics. For example, pulsar
scintillation studies had measured fluctuation spectrum of ISM over a
six-order-of-magnitude scale from $10^6$ to $10^{12}$ m, although the details
are still under debating (for evidence supporting Kolmogorov
spectrum see \citealt{ARS95}, for the deviations see \citealt{BRG99III}),

By carefully checking the pulsar dynamic spectrum, i.e. the pulsar
radio flux as the function of observing frequency and the time, one
may observe organised criss-cross structures. Such scintillation
phenomenon had been noted for nearly 30 years \citep{HWG85,CW86}.
Later, \citet{SMC01} discovered parabolic arc shape structures in
the secondary spectrum, i.e. the two dimensional Fourier transform
of dynamic spectrum \footnote{The arc like structure already appeared
in the work of \citet{CW86}, but later \citet{SMC01} drew attention to
the phenomena and provide the physical explanations.}.  The
scintillation arcs yielded additional insights into the scattering
process and provided an important perspective for the interstellar
medium, yet, the theories \citep{WMSZ04, CRS06} of scintillation
arcs had not only succeeded in explaining the phenomenon but also
provided quantitative models to infer the physical conditions of
ISM, e.g. the transverse velocity, the ISM screen distance, as well
as the ISM structures.

In this paper we focus on the observations of PSR B0355$+$54, for
which \citet{S06} had detected its scintillation arcs at 1.4 GHz.
We investigate the scintillation properties of PSR B0355$+$54 at a
higher frequency (2.25 GHz) using the Kunming 40-m telescope (KM40m).
In \S~\ref{sec:obs},  we introduce the setup of our observations.
The data analysis is in \S~\ref{sec:dan}. The discussions
and conclusions are made in \S~\ref{sec:condis}.

\section{Observations}
\label{sec:obs}

In the current paper observation of PSR B0355$+$54 was carried out
at KM40m radio telescope operated by \emph{Yunnan Astronomical
Observatory} (YNAO).  Being built in 2006 for the Chinese lunar-probe
mission, the telescope locates in the south west of China (${\rm
N} 25^\circ01'38''$, ${\rm E} 102^\circ47'45''$), approximately 15
kilometers away from a nearby city, Kunming.  The total collecting
area of KM40m is 1250 m$^2$. There is a room temperature S/X dual-band
circularly-polarised receiver installed for satellite tracking
purpose.  The system temperature of the receiver is 70 K at S-band.
The radio frequency signal is down converted to the intermediate
frequency, which has 300 MHz bandwidth.

In our observation, we recoded the IF signal with an 8-bit sampling
backend, \emph{pulsar digital filter bank 4} (DFB4) built
by the \emph{Australian National Telescope Facility} (ATNF).  We
captured the pulsar signal with the 512-channel configuration, each
of the channel has the width of 1$~$MHz. After integration with
sampling time of 64 $\mu$s, the audio data was folded with 512 bins
and 30 second sub-integration to form the data archive.

We performed 25 observations spreading across 60 days for PSR
B0355$+$54, i.e. observed from the end of January 2014 to the
beginning of April 2014. The length of observation varies from 30
to 120 minutes, depending on the telescope schedule. The telescope is close to
a city, this results in strong radio frequency interference (RFI). The RFIs left
60 to 130 MHz clean band in the original 300 MHz raw band.  The effective centre
of frequency is $2.25$ GHz. In each observation session, we had
checked the dynamic range of system and adjusted the voltage level
to keep the system in the linear regime.

\section{Data Analysis}
\label{sec:dan}

We apply three types of well-known methods  to study
the
scintillation process of PSR B0355+54, namely, the dynamic spectrum as in
\S~\ref{sec:dys}, the auto-correlation function (ACF) of dynamic
spectrum (\S~\ref{sec:acf}), and the secondary spectrum (\S~\ref{sec:secs}).

\subsection{Dynamic Spectrum}
\label{sec:dys}

The dynamic spectrum is a two dimensional presentation of radio
flux as a function of time and frequency. Due to the narrow pulse
(10\% for PSR B0355+54),  we gate the signal of each sub-integration.
We average the intensity with phase ranging from -0.1 to 0.1 centred
at the pulse peak to get the `on' flux. In a similar way, we calculate
the `off' flux by averaging intensity of alternative phases.
The pulse flux is then calculated as the difference between the
`on' and the `off' value.

\begin{figure}
		\includegraphics[width=3.5in]{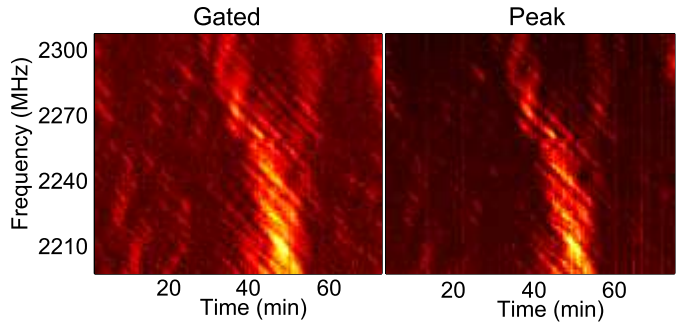}
		\caption{Comparison of dynamic spectra calculated using two different
		methods.  The left and the right panels are for the gated-flux method and
		the peak-flux method respectively. Both methods produce similar results.
		The gated-flux method recovers low-flux structures better than that of
		peak-flux technique. The data was taken on April 1st 2014 with KM40m.
		\label{fig:cp12}}
    \end{figure}

We have compared such gated-flux method with another well-known
technique, the peak-flux method, where instead of adopting the gated
average flux, the peak value is used.  Since the gated method integrates the
flux over the pulse phase, we
expect that it is more reliable, especially when the pulsar flux is low due
to the scintillation. The comparison between the two methods is given
in Figure~\ref{fig:cp12}, where it shows that the both methods
produce similar dynamic spectra, yet, as we expected,
the gated-flux scheme recovers the lower flux structures better
than what the peak-flux method does. We use the gated scheme in all
of following analysis.

Being close to the near-by city Kunming, KM40m is highly affected
by RFIs. There is no pulsar baseband recorder installed at KM40m at the time
when the observations were carried out. In this way, we could not use
voltage domain methods for RFI mitigation, e.g. the cyclic spectroscopy
\citep{WDV13} and the spectral kurtosis method \citep{NGL07}.
For RFIs with persistent frequencies, we replace the RFI-affected channels
 using the linearly interpolation from the adjacent channels.
We also interpolate in the time domain to remove spontaneous
wideband RFIs. Occasionally, broadband RFIs last for a few
minutes, and the resulted big data gap could not be repaired using
the interpolation. We then treat the data of each part separately
as different observations.

\begin{figure*}
\centering
\includegraphics[width=\textwidth, angle=0]{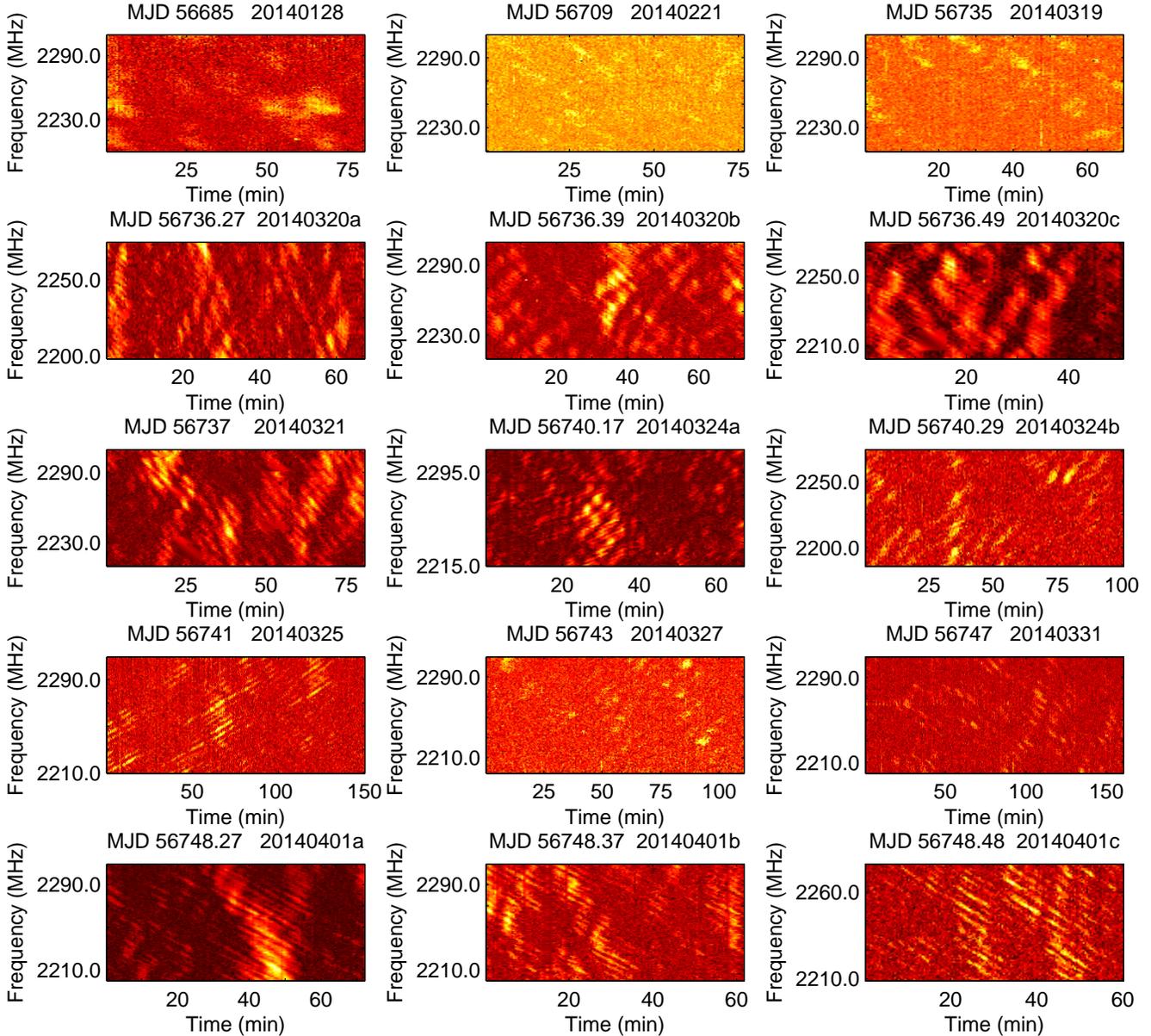}
\caption{Dynamic spectra of PSR B0355+54 at centre frequency 2.25 GHz taken from
January 28th 2014 to April 1st 2014. The epochs of observations are shown at the
top of panels. The horizontal and vertical axes correspond to the time and the
frequency,
respectively.
\label{fig:alldyn}}
\end{figure*}

In total, we measured 25 independent dynamic spectra for PSR B0355+54.
Fifteen of them have good signal-to-noise ratio, $S/N\ge 100$, and
these data show structured patterns as in Figure~\ref{fig:alldyn}.
The other 10 observations have lower $S/N$, and we could not found
clear structures after visual inspections. As one can see, these 15
high-$S/N$ dynamic spectra vary significantly as function of time.
The dynamic spectra showed scintles with roughly 50-MHz bandwidth
at beginning of January. The scintillation bandwidth reduced
significantly in March, and over merely one day (MJD 56735 to 56736),
the dynamic spectra developed into fringe-like structures.  Such
structures lasted for a month and turned into criss-cross structures
in the early April, while the scintillation bandwidth was continuously
reducing. Clearly, the scintillation parameters had changed significantly for
PSR B0355+54. In next
section, we provide quantitative analysis for such variations.

\subsection{The two-dimensional auto-correlation function}
\label{sec:acf}

To obtain the scintillation bandwidth and the decorrelation time-scale, we
compute
the two-dimensional (2-D) ACF of the dynamic spectrum \citep{C86, GRL94, BRG99I,
WYM08, Ball2014}, particularly, we follow the recipes from \citet{C86} and
\citet{WYM08}.  The 2-D ACF, $F(\Delta\nu,\Delta \tau)$, is calculated from
the pulsar flux, $S (\nu,t)$, as \begin{equation}
	F(\Delta\nu,\Delta\tau)= \sum_{\nu}\sum_{t} \Delta S(\nu,t)\Delta S(\nu+\Delta
	\nu,t+\Delta\tau)\,,
\label{eq:cor}
\end{equation}
where $\nu$ is the channel frequency, $t$ is the time, $\Delta \nu$ and
$\Delta\tau$ are correlation bandwidth and timescale respectively. The variation
of flux is defined as
$\Delta S(\nu,t)=S(\nu,t)-\overline{S(\nu, t)}$. We can normalise the 2-D ACF
using its value at the zero lag and the normalised ACF ($\rho$) is
\begin{equation}
	\rho(\triangle\emph{v},\Delta\tau)= F(\triangle\emph{v},\Delta\tau)/F(0,0)\,.
\label{eq:cor_norm}
\end{equation}

\begin{figure*}
\centering
\includegraphics[width=\textwidth, angle=0]{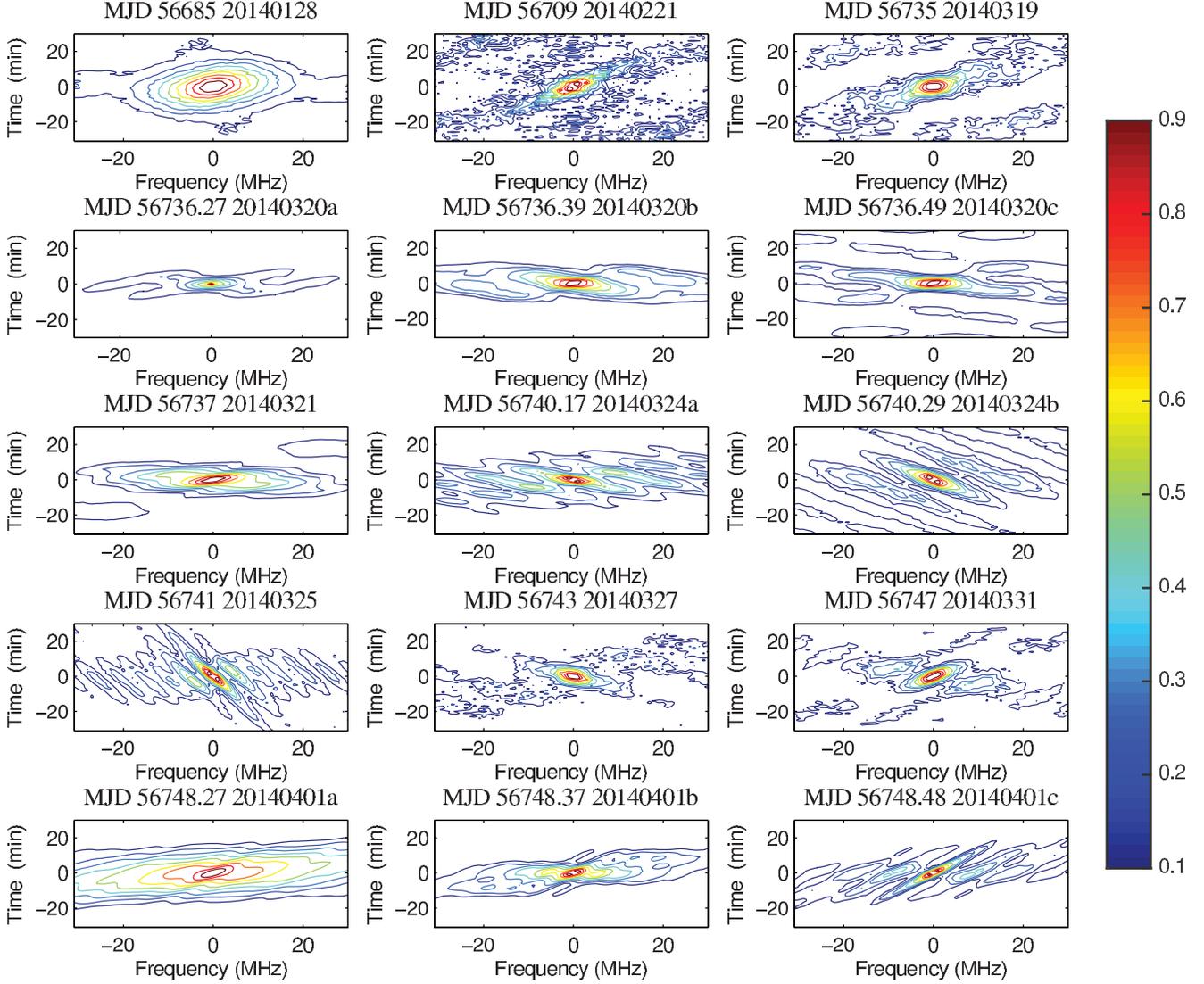}
\caption{Contour plots of normalised ACF of dynamic spectra shown in
Figure~\ref{fig:alldyn}.  The value of each contour is indicated using the
colour bar on the right side. \label{fig:allacf} }
\end{figure*}

\begin{figure*}
\centering
\includegraphics[width=\textwidth, angle=0]{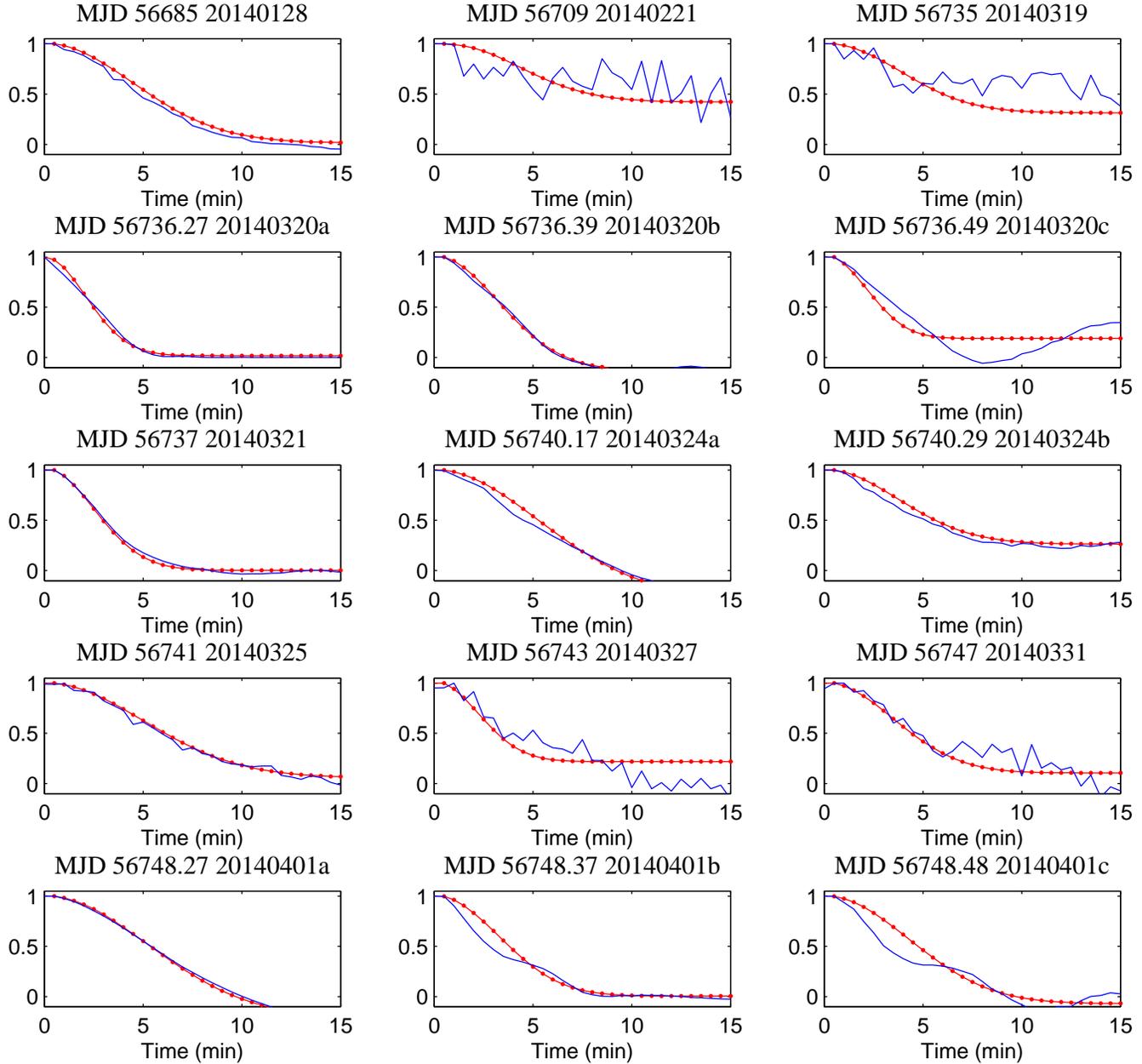}
\caption{One dimensional ACF in time, i.e. the $\Delta\tau_{\rm d}$-cross
section of the two-dimensional ACF. The solid line is the measured value, and
the dotted lines are from the best fitted two-dimensional elliptical Gaussian
function as explained in the main text.  \label{fig:alltacf}}
\end{figure*}

\begin{figure*}
\centering
\includegraphics[width=\textwidth, angle=0]{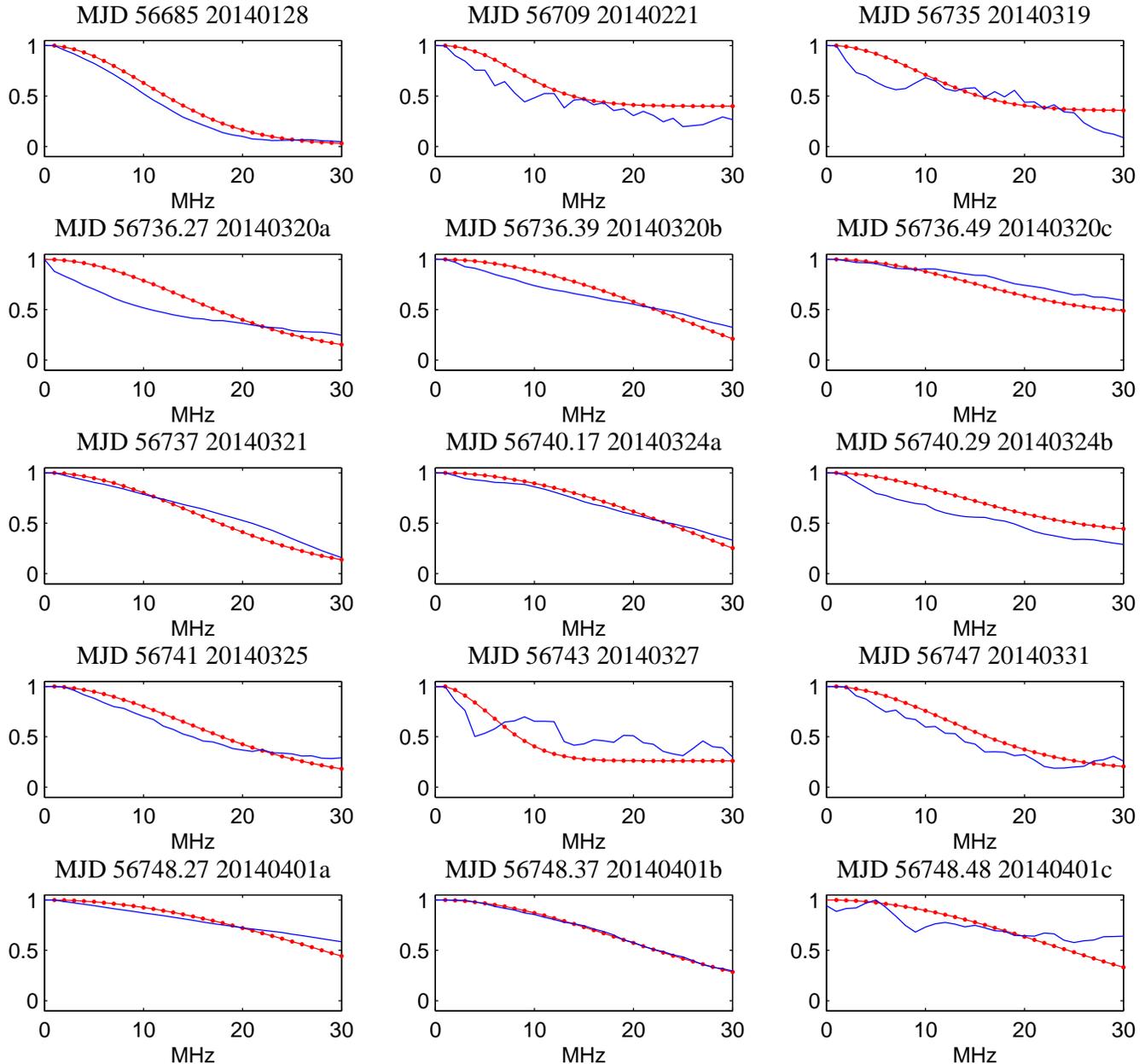}
\caption{Similar to Figure~\ref{fig:alltacf}, but the one dimensional ACF in
frequency.  The solid line is the measured value, and the dotted lines are from
the best fitted two-dimensional elliptical Gaussian function.
\label{fig:allfacf}}
\end{figure*}

The contour plots of the normalised ACFs are shown in
Figure~\ref{fig:allacf}, and the corresponding one-dimensional ACF
along the time and frequency axes are shown in
Figure~\ref{fig:alltacf}
and Figure~\ref{fig:allfacf}.

We use 2-D Gaussian fitting \citep{WYM08} to determine the scintillation
parameters
$\Delta\tau_{\rm d}$ and $\Delta \nu_{\rm d}$ according to the model that
\begin{equation}
	\rho_{\rm model}(\Delta\nu_{\rm d},\Delta\tau_{\rm d})=e^{-C_{1}\Delta\nu_{\rm
	d}^{2}-C_{2}\Delta\nu_{\rm d} \Delta\tau_{\rm d}-C_3\Delta\tau_{\rm d}^{2}}\,.
\end{equation}
The parameters $C_{1}$, $C_{2}$, and $C_{3}$ describe the size and the
orientation of elliptic of ACF. If we organised the coefficients $C_{1,2,3}$ in
to matrix $\rm \bf C$
\begin{equation}
	{\rm \bf C}=\left(\begin{tabular}{c c}
		$C_1$ & ${C_2}/{2}$ \\
		${C_2}/{2}$ & $C_3$ \end{tabular}\right)\,.
	\label{eq:elipmatrix}
\end{equation}
the orientation and size of the elliptic can be found using eigenvector and
eigen values of matrix $\rm \bf C$ as shown in \citet{LGY12}.

Such the Gaussian fitting is empirical, e.g. the Kolmogorov turbulent spectrum
predicts power-law correlation instead of Gaussian correlation \citep{RCN14}.
However, in order to compare the results of previous works, we keep such
Gaussian fitting convention. The discrimination of the correlation function
types is left for the future investigation.

We infer the values of $C_1, C_2$ and $C_3$ using the two dimensional $\chi^2$
fitting which minimises
\begin{equation}
	\chi^2=\sum_{\Delta\nu_{\rm d}, \Delta\tau_{\rm d}} \left[\rho(\Delta \nu_{\rm
	d}, \Delta\tau_{\rm d})-\rho_{\rm model}(\Delta \nu_{\rm d}, \Delta\tau_{\rm
	d})\right]^2\,.
\end{equation}
Following the usual conventions
\citep{BRG99I, WYM08}, the scintillation time scale $\Delta\tau_{\rm d}$ is the
half width
of time lag producing $\rho=1/e$, and the decorrelation bandwidth scale
$\Delta \nu_{\rm d}$ is the half-width of frequency lag giving $\rho=1/2$.
The scintillation parameters $\Delta\nu_{d}$ and $\Delta\tau_{d}$ are calculated
as
\begin{eqnarray}
\Delta\nu_{d}&=&\sqrt{\ln2/C_{1}}\,, \\
\Delta\tau_{d}&=&1/\sqrt{C_{3}}\,.
\end{eqnarray}
The error of estimated parameters come from two major sources,  {\bf i}) the
statistical error as computed from the $\chi^2$ fitting procedure
\citep{NR3}, and {\bf ii}) the error due to a finite number of bright scintles
in the
given data \citep{C86}.  The fractional error of the second type is \citep{C86,
Ball2014} $\sigma\simeq f^{-1/2} (\Delta\tau_{\rm
d}/T)^{1/2} (\Delta \nu_{\rm d}/{\rm BW})^{1/2}$, where $T$ and ${\rm BW}$ are
the data duration,
and bandwidth. The filling factor, $f$, is the ratio between the area of
bright scintles and the area for a single feature in the dynamic spectrum, which
is
choose as $f=0.2$ in the current paper.

\begin{figure}
\centering
\includegraphics[width=3.5in, angle=0]{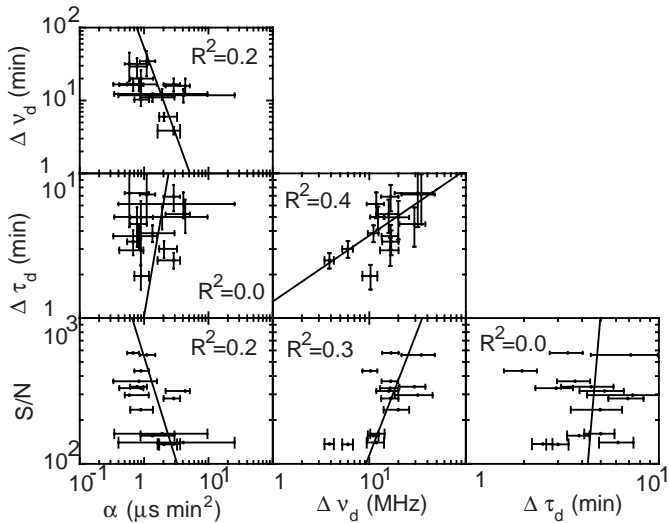}
\caption{The correlation between $S/N$, scintillation time scale, scintillation
bandwidth, and arc curvature with values from Table.~\ref{tab:nut}. The solid
line are the power-law function fitting.  The value of $R^2$ in each
panel is the coefficient of determination.  Except $S/N-\Delta\tau_{\rm d}$ and
$S/N-\Delta\tau_{\rm d}$ parameter pairs, the other parameter pairs show
marginal correlation.
Using the $\Delta\tau_{\rm d}-\Delta\nu_{\rm d}$ correlation, we can measure the
transverse velocity of PSR B0355$+$54, that $V/\sqrt{\eta}=310\pm 100$ km/s,
assuming pulsar distance $D=1.0$ kpc \citep{CCL03}.  \label{fig:nutc} }
	 \end{figure}

We summarise the measured values of $\Delta\nu_{\rm d}$ and $\Delta\tau_{\rm
d}$ in Table.~\ref{tab:nut} and Figure~\ref{fig:nutc}. Our
measurements agree with the predication of the Galactic free electron
density model NE2001 \citep{CL02}, where the predicted scintillation
decorrelation bandwidth and timescale of PSR
B0355+54 are 25 MHz and 390~s at the central frequency of 2.25 GHz, assuming a
transverse velocity of 100 km/s.

\begin{table*}
	\caption{Scintillation parameters for each epoch.
		The sup-script $a, b$ and $c$ indicate that we have split the session due to
		the RFI reason. $\Delta\tau_{\rm d}$ and $\Delta\nu_{\rm d}$ are the
		scattering timescale and decorrelation bandwidth respectively. The $\alpha$
		and $w$ are the curvature and width of the scintillation arcs. $d$ is the
		scattering screen-Earth distance infered from the arc curvature, where we
		had neglected the annual variations of Earth velocity due to the limited
		time span and used reference value of $V_{\rm eff, \perp}=100$ km/s. The
		detailed definitions are delayed in Section ~\ref{sec:secs}. For
		numbers with two values of error, the first error is the statistical error,
		while the second is the finite-sample error as we had explained.  We quote
		the logarithmic value of arc curvature, because
		the confidence level contours are more symmetric in logarithmic scale than
		linear as shown in Figure~\ref{fig:aw}.
		\label{tab:nut}
	}
	\begin{center}
	\begin{tabular}{lcccccccc}
		\hline \hline
	Epoch   & Duration& Bandwidth & $\Delta\tau_{d}$ & $\Delta \nu_{\rm d}$ &
	$\alpha$ & $w$ & $d_{\rm eff}$ & $S/N$ \\
	MJD  & min & MHz & min & MHz & $\log(\mu$s$\cdot${\rm min}$^2)$ &
	$10^{-2}\mu$s & kpc & \\
\hline
56685&80&110&6.1$\pm$0.2$\pm$1.2 &11.8$\pm$0.2$\pm$2.4      &$0.6^{+0.8}_{-1.0}$  & $9^{+16}_{-7}$   & 0.2&140\\
56709&76&100&2.5$\pm$0.2$\pm$0.2 &3.9$\pm$0.3 $\pm$0.3      &$0.5^{+0.1}_{-0.3}$  & $9^{+6}_{-5}$    & 0.2&137\\
56735&70&100&3.0$\pm$0.2$\pm$0.3 &6.0$\pm$0.3 $\pm$0.7      &$0.3^{+0.2}_{-0.1}$  & $8^{+2}_{-4}$    & 0.1&136\\
56736$^a$&84&61&2.9$\pm$0.2$\pm$0.6 &16.5$\pm$0.2$\pm$3.6   &$-0.1^{+0.05}_{-0.3}$  & $7^{+18}_{-3}$ & 0.03&331\\
56736$^b$&73&100&4.5$\pm$0.1 $\pm$1.3 &29$\pm$1.0$\pm$8.7   &$-0.1^{+0.2}_{-0.1}$  & $8^{+6}_{-3}$   & 0.03&338\\
56736$^c$&51&67&2.0$\pm$0.2 $\pm$0.3 &10.3$\pm$0.3$\pm$1.8  &$-0.1^{+0.1}_{-0.1}$  & $9^{+6}_{-3}$   & 0.03&434\\
56737&80&100&3.4$\pm$0.1 $\pm$0.6 &16.9$\pm$0.5$\pm$3.2     &$-0.2^{+0.1}_{-0.1}$  & $11^{+3}_{-5}$  & 0.03&576\\
56740$^a$&67&100&7.3$\pm$0.3 $\pm$3.0 &32$\pm$1.0$\pm$13    &$-0.2^{+0.3}_{-0.1}$  & $21^{+7}_{-9}$  & 0.03&296\\
56740$^b$&100&71&5.2$\pm$0.5 $\pm$1.2 &16$\pm$1.0$\pm$3.8   &$0.6^{+0.1}_{-0.3}$  & $13^{+8}_{-5}$   & 0.2&316\\
56741&150&100&6.9$\pm$0.5$\pm$1.3 &16.6$\pm$0.9$\pm$3.2     &$0.5^{+0.1}_{-0.2}$  & $16^{+5}_{-6}$   & 0.2&281\\
56743&110&115&5.0$\pm$0.4$\pm$0.8 &12.3$\pm$0.8$\pm$1.9     &$0.3^{+0.7}_{-0.7}$  & $12^{+6}_{-9}$   & 0.1&161\\
56747&160&110&3.9$\pm$0.3$\pm$0.4 &11.1$\pm$0.7$\pm$1.2     &$0.1^{+0.3}_{-0.2}$  & $13^{+7}_{-7}$   & 0.06&156\\
56748$^a$&80&110&7.2$\pm$0.1$\pm$2.7 &34.6$\pm$0.7$\pm$13   &$0.0^{+0.1}_{-0.1}$  & $14^{+8}_{-6}$   & 0.05&558\\
56748$^b$&77&110&3.7$\pm$0.1$\pm$0.7 &16.7$\pm$0.6$\pm$3.2  &$-0.1^{+0.3}_{-0.4}$  & $17^{+6}_{-10}$ & 0.03&368\\
56748$^c$&75&81&5.0$\pm$0.3$\pm$1.4 &20$\pm$1.0$\pm$5.7     &$-0.1^{+0.2}_{-0.2}$  & $18^{+9}_{-6}$  &0.03 &235\\

\hline \hline\end{tabular}
\end{center}
\end{table*}

For the thin-screen model of diffractive scintillation, the scattering time
scale and the bandwidth follow a simple relation that \citep{GRL94, WYM08}
\begin{equation}
	\Delta\tau_{\rm d}=	6.4\times 10^{2} \eta^{\frac{1}{2}} \left(\frac{D}{\rm
	kpc} \right)^{\frac{1}{2}} \left(\frac{\Delta \nu_{\rm d}}{\rm MHz}
	\right)^{\frac{1}{2}} \left(\frac{V}{\rm km/s} \right)^{-1}
	\left(\frac{\nu}{\rm GHz}\right)^{-1}  {\rm min}\label{eq:taunu}\,,
\end{equation}
where the $D$ and $V$ are the pulsar scintillation distance and velocity
respectively,  $\eta$ is the ratio between the screen-to-observer distance and
screen-to-pulsar distance. One can find the measured $\Delta\tau_{\rm d}-\Delta
\nu_{\rm d}$ relation in
Figure~\ref{fig:nutc}. After fixing the pulsar distance to the interferometry
distance $1.0$ kpc
\citep{CCL03}, we can fit the data using Equation.~\ref{eq:taunu} to get the
pulsar scattering velocity $V/\sqrt{\eta}=340\pm 130$ km/s. Comparing to the
interferometry velocity of 61 km/s \citep{CCL03}, one would require $\eta \simeq
0.01-0.1$, i.e. a rather nearby scattering screen locates $0.01-0.1$ kpc away
from the Earth in the pulsar direction.

\subsection{Secondary Spectrum}
\label{sec:secs}

The secondary spectrum $\cal S$ is the two-dimensional Fourier
transform of a dynamic spectrum.  Parabolic shaped structures (i.e.
the scintillation arcs) are expected in the secondary spectrum
\citep{WMSZ04, CRS06} under two conditions, 1) the scattering is
anisotropic that the interference is dominated by only a few
scattering directions; and 2) there is enough time-frequency
resolution and high $S/N$.

At lower frequencies, \citet{S06} had already detected scintillation
arcs for the PSR B0355+54, indicating the anisotropic scatterings.
The flux of PSR B0355+54 is 23 mJy at L-band, and its scattering
timescale and bandwidth are at the level of a few minutes and tens
of MHz.  All these make PSR B0355+54 a potential target to search for the
scintillation arcs at the 2.25 GHz.

\begin{figure*}
	 \centering
   \includegraphics[width=\textwidth, angle=0]{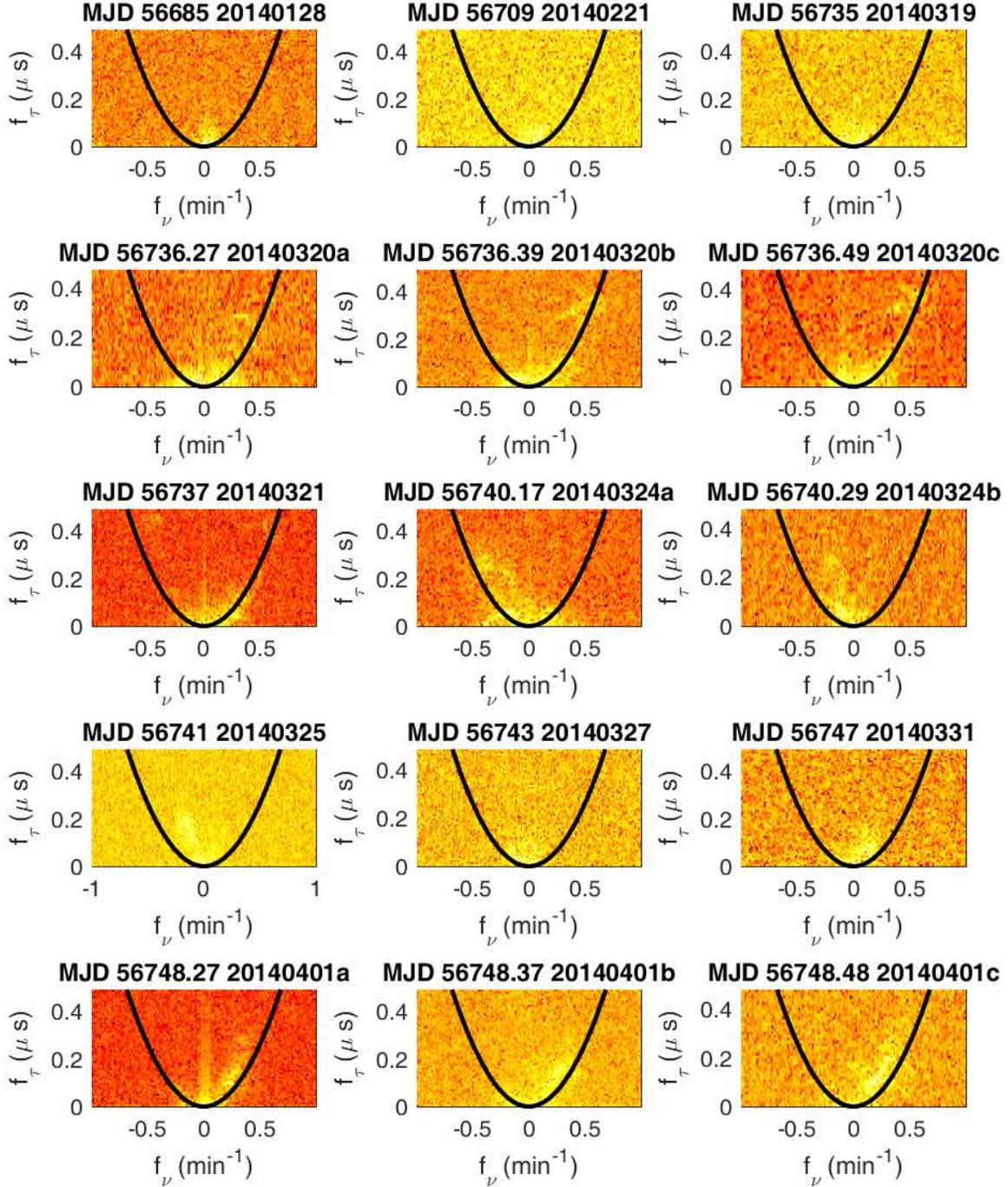}
	 \caption{Secondary spectra for the PSR B0355+54 displayed using a logarithmic
	 colour scale, i.e. $\log {\cal S}$. The x-axis and y-axis are the conjugate
	 time ($f_{\rm t}$) and conjugate
	 frequency ($f_{\rm \nu}$). The parabolic shaped
	 scintillation arcs can be spotted for most of the epochs. The black solid
	 parabolic curve indicates the weighted average curvature $\bar \alpha$, see
	 Equation~\ref{eq:bara} and related discussions for the details.
	 \label{fig:senscat}}
 \end{figure*}

The secondary spectra of our observations are plotted in
Figure~\ref{fig:senscat}. In multiple epochs, the
scintillation arcs are clearly visible. For epochs of March 20th
2014, March 24th 2014, and probably 1st April 2014, there may be
hints for the inverted arclets, but the limited $S/N$ and spectral
resolution prevent us from studying the details.

\citet{WMSZ04} and \citet{CRS06} had explained the physics behind
scintillation arcs. We have prepared a pictorial illustration
in Appendix~\ref{app:pic} to explain the relation between the scintillation
arc and the intensity distribution of scattered radiation. Basically,
the scintillation arcs come from the coherent interference
between a limited number of signal propagating paths.

For the most simple case, where the radiation gets scattered into
only two directions,  ${\boldsymbol \theta}_1$ and ${\boldsymbol
\theta}_2$, the intensities of the secondary spectrum will
distribute around those conjugate frequency ($f_{\nu}$) and conjugate
time ($f_{\rm t}$) with \begin{eqnarray}
 f_{v}&=&\frac{D\eta}{2c}\left (\theta_{2}^{2}-\theta_{1}^{2}\right)\,,
 \label{eq:fnu_par} \\
	f_{t}&=&\frac{1+\eta}{\lambda}({\boldsymbol{\theta}}_{2}-{\boldsymbol{
	\theta}}_{1})\cdot {\rm \bf V}_{\rm eff}\,. \label{eq:ft_par}
\end{eqnarray} Here, $\lambda$ is the observing wavelength, the
effective perpendicular velocity (${\rm \bf V}_{\rm eff, \perp}$)
is defined as \begin{equation}
	{\rm \bf V}_{\rm eff, \perp} =  {\rm \bf V}_{\rm psr,
	\perp}\frac{\eta}{1+\eta}+ {\rm \bf V}_{\rm obs,
	\perp}\frac{1}{1+\eta} - {\rm \bf V}_{\rm  screen, \perp}\,,
\end{equation} and ${\rm \bf V}_{\rm psr}, \perp$, ${\rm \bf V}_{\rm
obs, \perp}$, and ${\rm \bf V}_{\rm screen, \perp}$ are the transverse
velocity of pulsar, observer, and screen respectively \citep{CR98}.

From Equation.~\ref{eq:fnu_par} and \ref{eq:ft_par}, one can see that the
$f_{\rm \nu}=\alpha f_{\rm t}^2$, if the major
scattered intensities are at the origin ($\boldsymbol{\theta}_1=0$) and along
a straight line passing through the origin, i.e. $\boldsymbol{\theta}_2=\theta
{\rm \bf n}$ with $\rm\bf n$ being
a unit vector in the scattering screen. In such a scenario, the intensity
distribution in the secondary spectrum will be the parabolic arc with a
curvature of
 \begin{equation}
	\alpha=\frac{D \eta \lambda^2}{2c ({\rm \bf V}_{\rm eff, \perp}\cdot {\rm \bf
	n})^2 (1+\eta)^2} \,.
	\label{eq:curvature}
\end{equation}

\begin{figure}
\centering
\includegraphics[width=3.5in]{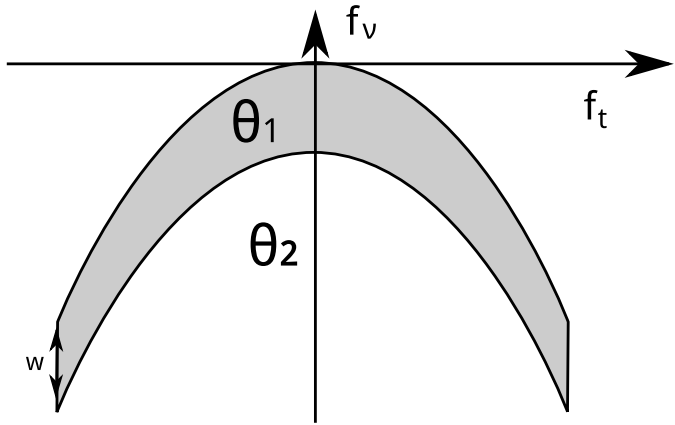}
\caption{The parameter space $\Theta_1$ and $\Theta_2$ for
Equation.~\ref{eq:hough}. The $\Theta_1$ is the parameter space spanned by the
arc with curvature of $\alpha$ passing the origin and the arc parallely shifted
by $w$ along
$f_{\rm \nu}$ axis, i.e. the parameter space in the parabolic belt with width of
$w$.  The statistics $ S'$ is defined as the difference between the reduced
$\chi^2$ of
$\log \cal S$ in the region $\Theta_1$ and $\Theta_2$.
\label{fig:int} }
\end{figure}

As shown in Figure~\ref{fig:senscat}, our observed scintillation arcs vary over
the 90 days, particularly, the curvature of arcs and distribution of intensity
along the arc change temporally.  Due to the limited $S/N$, we can not study the
arc variation by tracing
the arclets as in \citet{S06} or \citet{TR07}. Alternatively, we
designed a statistics $S'$ to do so. The statistics is similar to the
generalised Hough
transformation \citep{B81}.  We
re-parameterise the parameter space of $f_{\rm \nu}$ and $f_{\rm t}$ using
the other two parameters, the parabolic width $w$ and arc curvature
$\alpha$.  The transformation is defined as the difference between the reduced
$\chi^2$ of secondary spectrum in the two complementary regions $\Theta_1$ and
$\Theta_2$. As illustrated in Figure~\ref{fig:int}, the region $\Theta_1$
contains the given arc and the region $\Theta_2$ does not.

The statistics $S'$ is defined as \begin{equation}
	{S'}=\sum_{ \{ f_{\rm \nu} f_{\rm t} \}\in \Theta_{\rm 1} }  \frac{\log
	{\cal 	S}}{N_{\Theta_1} \sigma_{\log\cal S}} - \sum_{ \{ f_{\rm \nu} f_{\rm
	t} \}\in \Theta_{\rm 2} }    \frac{\log {\cal 	S}}{ N_{\Theta_2} \sigma_{\log
	\cal S}} \,,
	\label{eq:hough}
\end{equation}
where the region $\Theta_{1}$ in the $f_{\rm t}-f_{\rm \nu}$ space covers the
parabolic arc with the
curvature of $\alpha$ and its width $w$ along the positive $f_{\rm \nu}$
direction, as illustrated in Figure~\ref{fig:int}. The complementary region
$\Theta_2$ covers the rest of $f_{\rm t}-f_{\rm \nu}$ space\footnote{Due to the
DC term, we removed the region of $f_{\rm t}=0$ and $f_{\rm \nu}=0$
from both $\Theta_1$ and $\Theta_2$.}.  We denote the standard deviation of
$\log S$ as
$\sigma_{\log \cal S}$. The number of data points in the two
regions $\Theta_1$ and $\Theta_2$ are denoted as $N_{\Theta_1}$ and
$N_{\Theta_2}$ respectively.

The statistics $S'$ is defined using logarithm of secondary spectrum, because
the probability
distribution of $S$ is very close to the log-normal distribution\footnote{ The
Fourier transform ($S$) of Gaussian signal follows a two-degree-of-freedom
$\chi^2$ distribution, i.e. the distribution function of the $S$ is
$f(S)\propto e^{-S}$.  The distribution of $y\equiv\log S$ is $f(y)\propto
e^{-e^y+y}$ according to random variable transformation. One gets $f(y)\propto e^{-y^2/2}$, when $|y|<1$, i.e. when $S$
does not vary by orders of magnitudes, i.e. $\log S$ behaves approximately as
a Gaussian distribution.}.
The statistics $S'$, therefore, describes the
significance of the difference between the mean of $\log S$ in the regions
$\Theta_{1}$ and $\Theta_2$. The distribution of $S'$ under the null hypothesis
that
there is no arc-like structure in the secondary spectrum follows the Student's
$t$-distribution, and we can use the $t$-test to compute the confidence level of
$w$ and $\alpha$.

With $S'$, we can measure the arc curvature and the direction of scattering, where the arc width $w$
is used to infer scattering angle $\theta_{\rm y}$ as illustrated in
Figure~\ref{fig:cartoon} of Appendix.~\ref{app:pic} that \begin{equation}
	\theta_{\rm y}=\sqrt{\frac{2 c w}{D \eta}}\,.\label{eq:tw}
\end{equation}

{We note that a recent work \citep{BOT16} is similar in analysing
the secondary spectra. There are differences between \citealt{BOT16}
and the current work.  Firstly, the statistics are chosen differently.
\citet{BOT16} integrated scintillation power along the path of
parabolic arc with a fixed pixel width, then measured the arc
curvature.  We rely on the statistics $\cal S'$ to perform the
parameter inference. The statistics $\cal S'$ is the likelihood
ratio test and it is also the matched filter to detect power difference
\citep{DR68}. Secondly, we perform statistical inference simultaneously
on the two parameters, $w$ and $\alpha$, while in \citet{BOT16} $w$
is fixed. As the two parameters show clear correlation,
we prefer to use the current two-parameter approach to get reliable
error estimation (see more discussion in \citet{NR3}).

The measured $S'$ as a function of $\alpha$ and $w$ are shown in
Figure~\ref{fig:aw}. As one can see that the most of the spectral
energy concentrates around a particular value of $\alpha$ and $w$. The value of width
$w$ is about $0.1 {\rm \mu s}$. According to Equation~\ref{eq:tw}, the
corresponding the angular scale of scattered radiation
is 0.2 mas.
We can also derive the effective scattering screen distance from the arc
curvature $\alpha$
via $d_{\rm eff}=2\alpha c {V}_{\rm eff, \perp}^2(1+\eta)/\lambda^2$ (see
Eq.~\ref{eq:curvature}).
The effective scattering scattering distance is defined as
	\begin{equation}
		d_{\rm eff}\equiv\frac{D\eta}{ (1+\eta) \cos^2 \theta_{\rm \perp}}\,,
		\label{eq:diseff}
	\end{equation}
	where the angle $\theta_{\rm \perp}$ is the angle between the scattering position $\rm\bf n$ and
	the transverse	effective velocity ${\rm \bf V}_{\rm eff, \perp}$. We list the
	measured $d_{\rm eff}$ in Table.~\ref{tab:nut}, which agree with the distance
	estimation from the  $\Delta\tau_{\rm d}-\Delta \nu_{\rm d}$ relations.

\begin{figure*}
\centering
\includegraphics[width=\textwidth, angle=0]{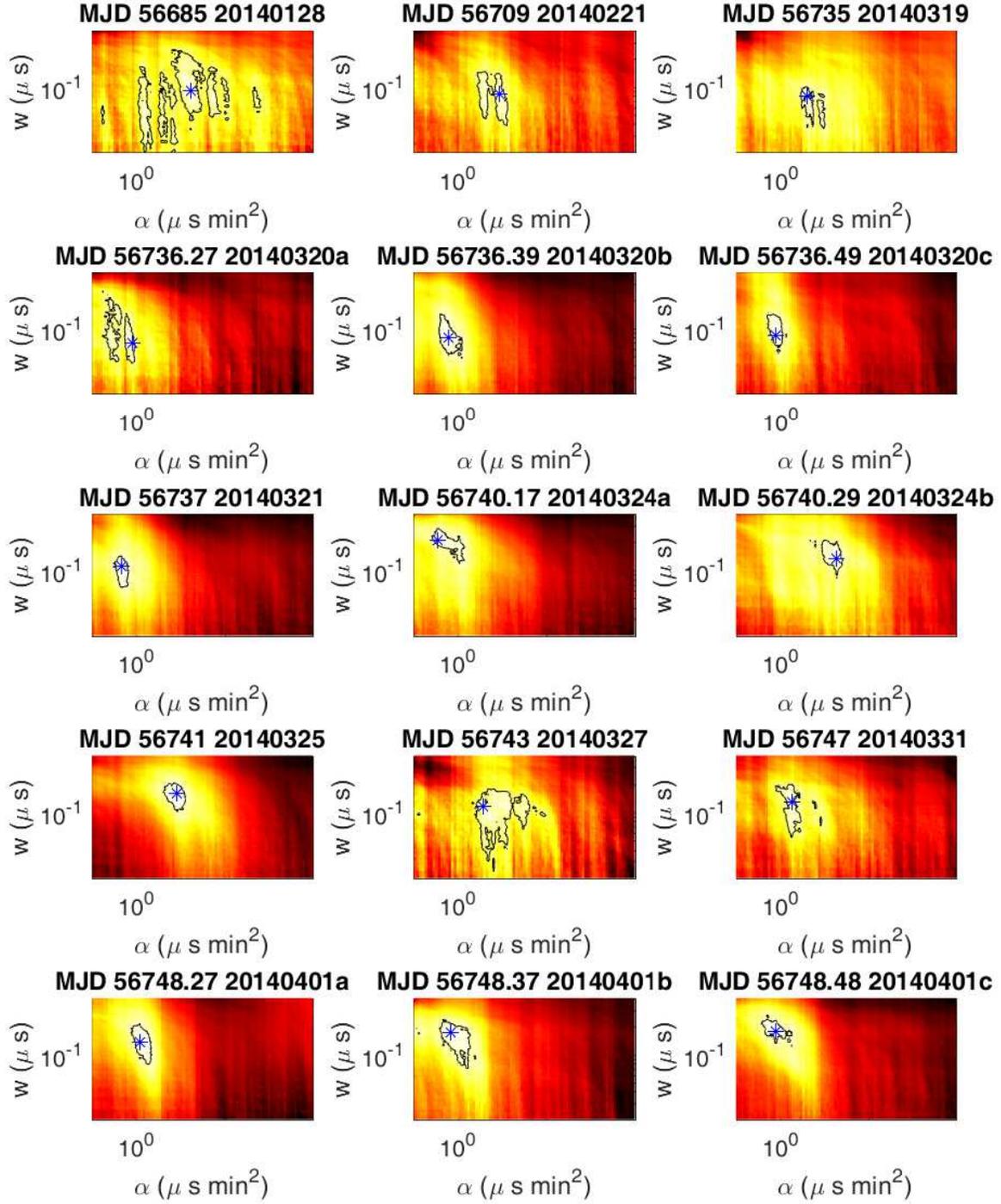}
\caption{The transformed secondary spectra $\cal S'$ as function of curvature
$\alpha$ and width $w$ according to Eq.~\ref{eq:hough}.  The black contours are
for `1-$\sigma$' confidence level, and `*' symbol indicates the most probable
values for $\alpha$ and $w$.
\label{fig:aw}}
\end{figure*}

}

We use the  $\chi^2$ test to check if the arc curvature is varying. Here the
null hypothesis $H_{0}$ and positive hypothesis $H_1$ are
\begin{equation}
	\left\{ \begin{array}{l}
		\text{$H_{0}$: arc curvature is a constant,} \nonumber \\
		\text{$H_{1}$: arc curvature is varying.}
	\end{array}\right.
\end{equation}
Under the $H_{0}$, the statistics $\chi^2$ is defined as \begin{equation}
	\chi^2=\sum_{i} \left(\frac{\alpha_i-\bar{\alpha}}{\sigma_{\alpha, i}}\right)^2\,
	\label{eq:chi2tst}
\end{equation}
which follows the $\chi^2$ distribution with degree of freedom of $N-1$. Here, $i$
is the index of observation session and $N$ is the total number of data points.
The weighted average value $\bar{\alpha}$ is defined as \begin{equation}
	\bar{\alpha}=\frac{\sum_{i} \alpha_{i} \sigma_{\alpha, i}^{-2}}{\sum_{i}
	\sigma_{\alpha, i}^{-2}}\,, \label{eq:bara}
\end{equation}
where $\sigma_{\alpha,i}$ is the error of $\alpha_i$

Since the errorbars of data points are asymmetric, the error
$\sigma_{\rm \alpha_{i}}$ is chosen accordingly depending on whether
curvature $\alpha_{i}$ is larger or smaller
than $\bar{\alpha}$.  Using the values in Table.~\ref{tab:nut}, we
get $\chi^2=40$. The corresponding P-value is $\le2\times10^{-4}$ under
$H_{0}$,
i.e. we can rule out the hypothesis that the arc curvature is a constant with
the probability of making mistake no more than $2\times10^{-4}$ (more confident
than the `3.5-$\sigma$'). From the transformed secondary
spectra, one can visually see how the arc curvature varies
epoch-to-epoch. In Figure~\ref{fig:senscat}, we also draw curves to indicate
the weighted mean $\bar{\alpha}$ for comparisons.

\begin{figure*}
\centering
\includegraphics[width=5in, angle=0]{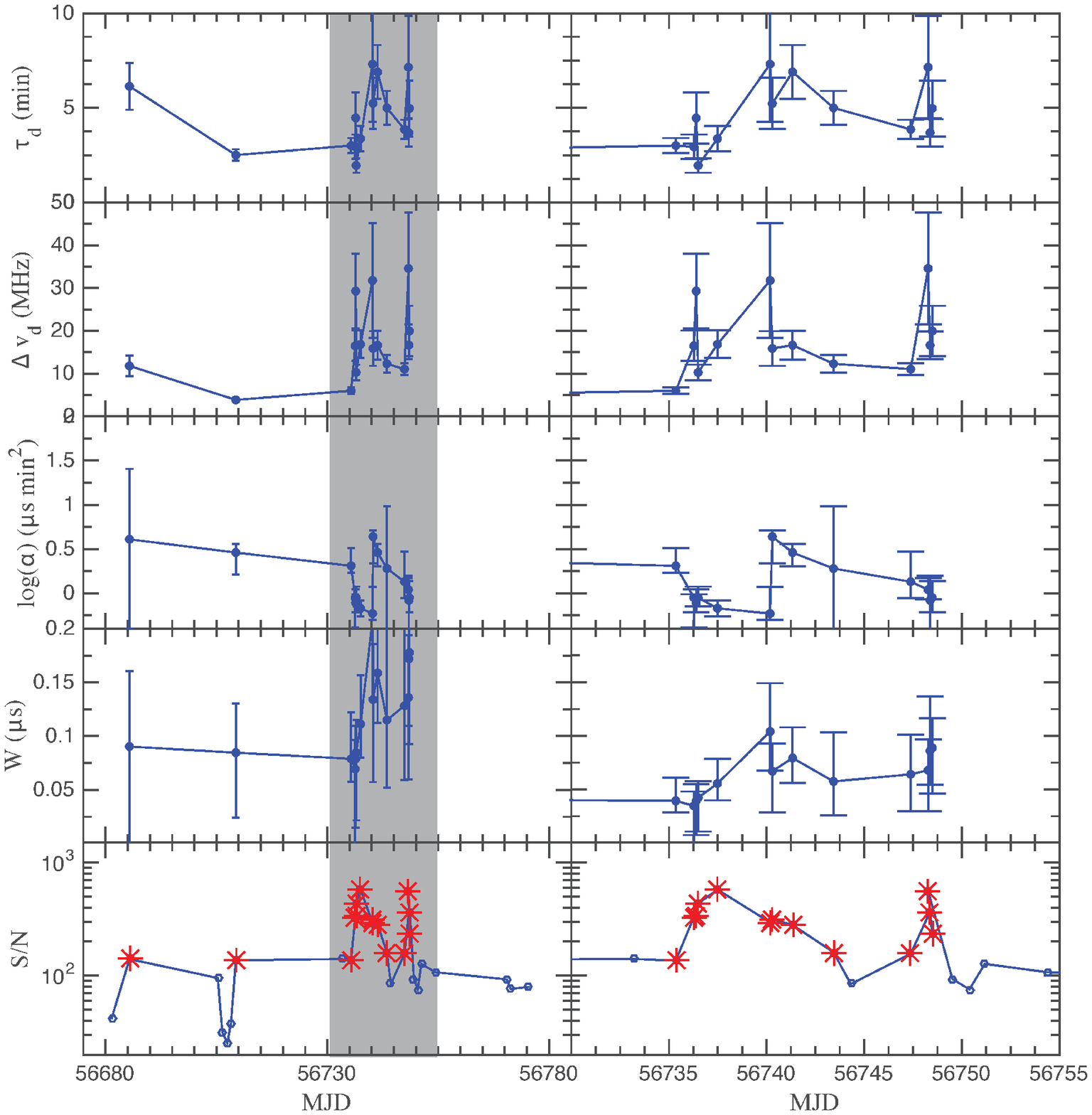}
\caption{The temporal variation of scattering timescale $\Delta\tau_{\rm d}$,
bandwidth $\Delta \nu_{\rm d}$, arc curvature $\alpha$, arc width $w$,
and $S/N$. The x-axis is the epoch of observation. One can see that all those
parameters are temporally variable except the arc width $w$. The right panels
show
the zoomed regions indicated by the grey shading in the left panels. In the
$S/N$ panel, the $*$ symbols indicate the observation epochs
showing scintillation arcs. The open circles are the epochs showing no arcs.
\label{fig:summary}}
\end{figure*}

As a summary, we collect all the temporal variation of scattering bandwidth,
scattering timescale, arc curvature, width, and $S/N$ in
Figure~\ref{fig:summary} and Table~\ref{tab:nut}. Due to the lack of noise
calibrator at the site, we can not convert the $S/N$ to the radio flux.
In this way,  $S/N$ is only for the
reference purpose. The $S/N$ is defined as the ratio between the
integrated intensity and noise intensity as computed by \textsc{PSRCHIVE}
\citep{HVM04}.

From Figure~\ref{fig:summary}, one can see that four scattering parameters vary temporally, i.e. the scattering bandwidth,
the scattering timescale, the arc curvature, and $S/N$. We have
also checked the correlation between $\Delta\tau_{\rm d}$, $\Delta
\nu_{\rm d}$, $S/N$, and arc curvature as in Figure~\ref{fig:nutc}.
We have used the coefficient of determination $R^2$, the fractional
reduction of total sum of squares \footnote{For the regression
problem, the coefficient of determination is defined as $R^2=1-\frac{\sum
(y_i-f_i)^2}{\sum (y_i-\bar{y})^2}$. Here $y_i$ is data values, and
function $f_i$ is the model value.}, to describe the correlation
between the four parameters. There is indication for correlation
between $\Delta \nu_{\rm d}$-$S/N$, and $\alpha$-$S/N$. There also
seems to be an anti-correlation between the arc curvature ($\alpha$)
and scattering bandwidth ($\Delta \nu_{\rm d}$).

\section{Discussions and Conclusions} \label{sec:condis}

In this paper, we had studied the scintillation of radio radiation
from PSR B0355+54 using the Kunming 40m radio telescope at 2.25
GHz. We had measured the dynamic spectra of PSR B0355+54 for 15
epochs. Using the ACF method, we have inferred the scattering
timescale and bandwidth, which agree with the predication of the
NE2001 model.  We used the generalised Hough transform to measure
and to study the curvature and scattering directions.

Although the arclets are not resolved in our measured secondary
spectra, we can see the arc are asymmetric with respects to the
$f_{\rm t}$ axis. We can see such $f_{\rm t}$-axis mirror symmetry
of scintillation arc varies on time scale of days,
which indicates that the asymmetry of interstellar scattering medium at the
0.1-AU scale, assuming velocity scale of $100$ km/s. This
agrees with results of \citet{BMG10}.

{ If we interpret our result in the thin screen scintillation model, then our
inferred scintillation screen will be very nearby (0.01-0.1 kpc). By comparing to
the geometry of local ISM structure of \citet{BG02}, our scattering screen seems
to be at the edge of the Local Bubble. However, due to the complexity in the
Local bubble and Perseus cloud \citep{LVV14, YM17}, it is still premature to
drew conclusions.}

A feature can be noted in both Figure~\ref{fig:senscat},~\ref{fig:aw},
and~\ref{fig:summary}, that the arc curvature varies on a rather
short timescale of days. For example, from March 24th 2014 (MJD
56740) to March 25th 2014 (MJD 56741), the arc curvature increased
from $\log\alpha=-0.2 \log {\rm \mu s\cdot min^2}$ to $0.6 \log
{\rm \mu s\cdot min^2}$ and later decrease to $-0.1 \log {\rm \mu
s \cdot min^2}$  at April 1st 2014 (MJD 56748).  This is surprising,
because the curvature is thought to be determined by pulsar distance,
effective velocity, and screen position.  Pulsar distance clearly
can be regarded as a constant over 90 days. The earth velocity and
screen velocity should be comparable to the proper motion of PSR
B0355+54. Earth velocity mainly leads to an annual variation of arc
curvature \citep{S06}, and will not contribute to the day-timescale
variation we found here.  The reason of such short timescale variation
of curvature, thus, should be either 1) the variation of distance
ratio $\eta$ of scattered screen or 2) the angle between transverse
velocity and scattering direction, i.e. $\theta_{\rm \perp}$ in
Eq.~\ref{eq:diseff} or 3) only the inner edge the scintillation arc
is visible for epochs with higher $\alpha$. In the first scenario, scattering
screens with different
distance dominate the scattering at different epochs \citep{S06,
S07}. In the second scenario, pattern of the scattering structure
changes or different scintillas dominate the interference
\citep{BMG10}. In the third scenario, one requires the arclets
be highly asymmetric.  Our detection of
anti-correlation between arc curvature ($\alpha$) and scintillation
bandwidth $\Delta \nu_{\rm d}$ seems to prefer the first interpretation.
A dedicated pulsar
monitoring using interferometer \citep{BMG10, BJK15} may help to
resolve the three possibilities.

The resolution of our secondary spectrum is limited, in fact, by RFIs.  The
observation bandwidth is limited by the RFIs, so the  $f_{\rm \tau}$ resolution
is limited. Splitting data into different segmentation reduces data length, so
do the resolution for $f_{\rm \nu}$.

{It seems whether we can observe the scintillation arc is related
to the value of $S/N$. As shown in Figure~\ref{fig:summary},
we did not find scintillation arcs in epoch with $S/N$ slightly
below the threshold of $100$. In principle, whether one observes the
scintillation arc
depends not only on the $S/N$, but also the anisotropy of scattering.
We expected that future telescopes with higher sensitivity, e.g.
FAST\citep{Nan11}, QTT\citep{Wang17}, or SKA\citep{Han15} will be beneficial in
the
pulsar scintillation studies.}

\section*{Acknowledgments}
We thank Prof. Coles, W. A. and the
anonymous referee for very helpful discussions for the interpretation
of curvature variation. This work was supported by
NSFC U15311243, National Basic Research Program of China, 973 Program,
2015CB857101,
XDB23010200, 11373011, 11303093, U1431113,
11178001 and 11573008. KJL is supported by the Max-Plack Partner Group.

\label{lastpage}
\bibliographystyle{mn2e}
\bibliography{ms}
\clearpage
\appendix

\section{Pictorial introduction to the scintillation arc}
\label{app:pic}

\begin{figure}
\centering
\includegraphics[width=3.0in]{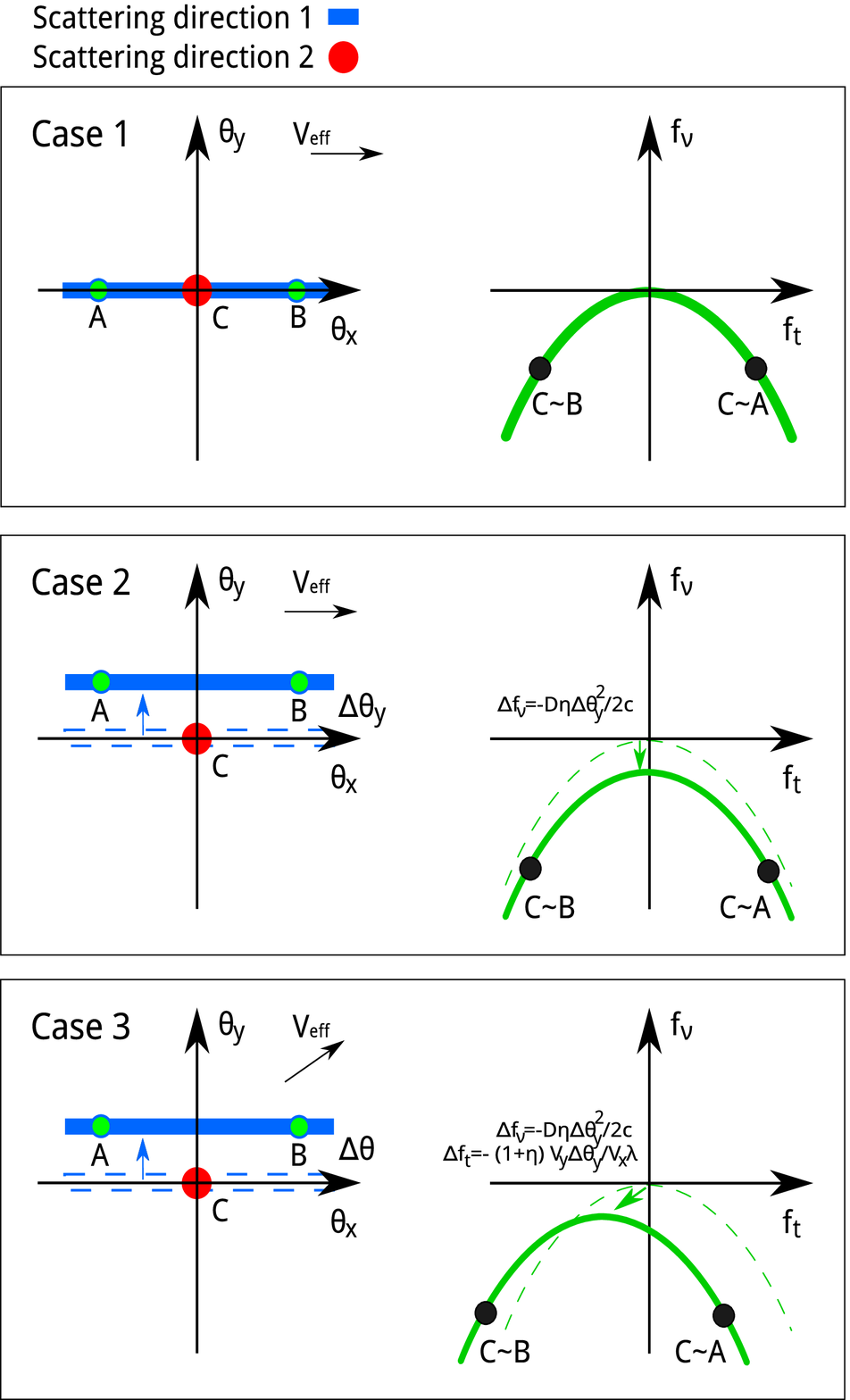}
\caption{Cartoon illustrations for the relations between scattered intensity
distribution and scintillation arcs. The left panels are the distribution of
scattered intensity. The right panels are the corresponding
scintillation arcs in the secondary spectra. Here the $V_{\rm eff}$ is the effective velocity, and the arrow indicates it direction.
For the most simple case, the scattered intensity is dominated by two major
components, a central component $C$, and a strip structure $A-B$.
The interference between point $A$ and $C$ is mapped into the
$C\sim A$, and similarly, the $B$-$A$ interference is mapped into $B\sim A$.
For the `Case 2' and `3', the value of arc shift in $f_{\rm t}$ and $f_{\rm \nu}$
parameter space are given in the figure.  \label{fig:cartoon}
}
\end{figure}

In
Figure~\ref{fig:cartoon}, we use cartoon figures to illustrate the relation
between the scintillation arc and the scattered intensity for a few
highly simplified scenarios. The details for the modeling can be found in
\citet{WMSZ04} and \citet{CRS06}.

Three cases are shown here. In all cases, the interference happens between a
localised central radiation and a strip-shaped scattered
radiation (see \citet{PL14} for a possible physical explanation of such 1-D
structure and \citet{CR10} for a more detailed simulation).  In the first
case, the effective velocity lies along the
elongate direction of strip-shaped radiations; and the strip structure overlays
with centre radiation. In the second case, the strip structure gets shifted
perpendicular
to the direction of effective velocity.  The third case is similar to the
second, except that the
effective velocity and elongating direction of radiation strip are misaligned.
The figure here show how such shifts and misalignments of radiation strip
structure lead to the corresponding shifts of parabolic scintillation arc in the
$f_{\nu}$-$f_{\rm t}$ parameter space.

Clearly, if the strip components have certain width along the $\theta_{\rm y}$
direction, the scintillation arc will gain finite width ($w$ defined in
Figure.~\ref{fig:int}), which is determined by the angular scale $\Delta
\theta_{\rm y}$ as shown in the figure.

\end{document}